\documentclass{aps2010}

\usepackage{graphicx,color}
\usepackage{amssymb}
\usepackage{amsbsy}
\usepackage{amsmath}
\usepackage{textcomp}

\def\bI{\mathbf{I}}
\def\bJ{\mathbf{J}}
\def\bR{\mathbf{R}}

\title{Performance of Polarimetric Beamformers for Phased Array Radio Telescopes}
\author[org1]{Marianna V. Ivashina}
\author[org2]{Stefan J. Wijnholds}
\author[org1]{Rob Maaskant}
\author[org3]{Karl F. Warnick}
\address[org1]{Chalmers University of Technology, SE-412 96 Gothenburg, Sweden. Email: ivashina@chalmers.se}
\address[org2]{ASTRON, Oude Hoogeveensedijk 4, NL-7991 PD, Dwingeloo, The Netherlands. Email: wijnholds@astron.nl}
\address[org3]{Brigham Young University, 459 Clyde Building, Provo, UT 84602, USA. Email: warnick@ee.byu.edu}

\begin{document}

\maketitleblock

\begin{abstract}
The results of four recently introduced beamforming schemes for phased array systems are discussed, each of which is capable to provide high sensitivity and accurate polarimetric performance of array-based radio telescopes. Ideally, a radio polarimeter should recover the actual polarization state of the celestial source, and thus compensate for unwanted polarization degradation effects which are intrinsic to the instrument. In this paper, we compare the proposed beamforming schemes through an example of a practical phased array system (APERTIF prototype) and demonstrate that the optimal beamformer, the max-SLNR beamformer, the eigenvector beamformer, and the bi-scalar beamformer are sensitivity equivalent but lead to different polarization state solutions, some of which are sub-optimal.
\end{abstract}

\section{Introduction}
The radio astronomy community has recently decided upon the science cases that future radio telescopes must enable, which directly translate into stringent requirements on the instrument's survey speed. These include beam sensitivity, size of the field-of-view (FoV), number of simultaneous beams, and frequency bandwidth, as well as on the polarization purity of the instrument. Conventional single-beam antennas are not capable to conform to these multi-beaming performance specifications. Research has therefore been focused on phased-array antenna concepts, since then the individual antenna array output signals can be weighted to realize multiple desired beams. Furthermore, array signal processing techniques can be used to exploit the degrees of freedom in the beamformer weights for increased beam sidelobe level control, interference rejection, and gain and noise level control of the beams (SNR)~\cite{Jeffs2008,Ivashina2011}. 

To date, the polarimetric response of the antenna array has not received much of attention and only one polarization is usually considered at a time. Further, as the radio polarimeter is essentially summing up vector element patterns, the polarization characteristics of the realized beam is weight dependent too. In turn, this enables us to create two beams in one direction forming an orthogonal polarimetric beam pair. Moreover, signal processing techniques can be developed to collectively optimize for the SNR and orthogonality of a polarimetric beam pair. This paper assesses the results for four different beamforming schemes that have recently been proposed: the optimal beamformer, the max-SLNR beamformer, the eigenvector beamformer, and the bi-scalar beamformer. The beam pair sensitivities are shown for the APERFIF prototype system~\cite{Ivashina2011}, for all beamforming scenarios. The polarimetric quality of each pair of nominally orthogonally polarized beams is expressed in terms of a Jones matrix. This formulation is used in a numerical model to study the performance of several possible algorithms for polarimetric array beamformer weight design.

\section{Polarimetric Beamforming}

\begin{figure}[h]
\centering
\def\svgwidth{0.8\textwidth}
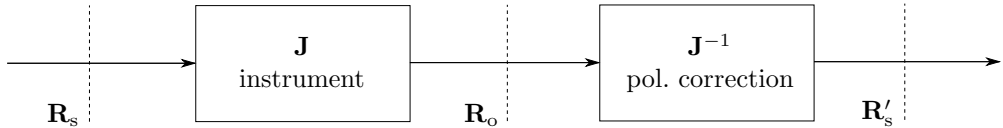
\caption{A system level view of a radio polarimeter. \label{fig:systemview}}
\end{figure}

Figure \ref{fig:systemview} provides a system level view of a polarimetric antenna system. The polarimetric properties of the source are defined by the $2 \times 2$ source covariance matrix $\bR_\mathrm{s}$, which gives the covariance of the source signal in two polarizations defined in a suitable coordinate system. The transfer function of our instrument, including antennas, receiver chains and a beamformer, can be described in the form of a $2 \times 2$ Jones matrix $\bJ$, which describes how the input voltages in two polarizations are transformed into two polarimetric output signals. If we correlate these output signals, we get the beamformer output covariance matrix $\bR_\mathrm{o}$. This is used as input for further processing, that ultimately produces a reconstructed source covariance matrix $\bR_\mathrm{s}^\prime$ that should ideally be proportional to $\bR_\mathrm{s}$. An ideal system does not require polarimetric correction, since it has $\bJ = \bI$, where $\bI$ denotes the identity matrix, i.e., it leaves the covariance matrix of the input signal unchanged and does not introduce so-called instrumental polarization. By choosing appropriate beamformer weights, the instrumental polarization introduced by the antennas and other analog electronics can be compensated for in the beamformer. In this paper, we compare the polarimetric and sensitivity performance of four proposed beamformers:

\begin{description}
\item[optimal beamformer] In this beamforming scheme, we assume that we know the response of the system to two reference signals with mutually perfectly orthogonal polarization. We then maximize the sensitivity under the constraint that $\bJ = \bI$ by minimizing the receiver output noise power~\cite{Warnick2010}.
\item[max-SLNR beamformer] The optimal beamformer imposes a hard constraint on the instrumental transfer function. By optimizing the signal-to-leakage-and-noise ratio, one can turn this into a soft constraint that does not suppress the leakage further than the noise level in the calibration measurement. This ensures that the noise in the calibration measurement is not over-interpreted.
\item[eigenvector beamformer] This more practical beamformer is based on the fact that an unpolarized signal can be regarded as a superposition of two uncorrelated orthogonally polarized signals. We can thus produce two orthogonal output signals by measuring the signal covariance matrix of the antenna outputs, which is a rank-2 matrix, and associating each beamformer output with one of the dominant eigenvectors \cite{Veidt2010}. If the phased array feed consists of two sets of single polarization elements, we can make an approximate polarimetric correction by exploiting the intrinsic polarization of each set of elements.
\item[bi-scalar beamformer] This beamformer assumes that the array consists of two sets of single polarization elements and simply applies max-SNR beamforming to these sets separately to obtain the two beamformer outputs.
\end{description}

\section{Numerical Results}

The following results pertain to the APERTIF prototype system which includes a phased array feed consisting of 144 tapered slot antenna elements, 144 low noise amplifiers and an active (digital) beamforming network. The entire receiving system has been modeled by an enhanced version of the method of moments used as the electromagnetic simulation software, combined with a microwave circuit simulator (CAESAR), and the high frequency electromagnetic solver GRASP for modeling the reflector antenna~\cite{IvashinaICEAA2010,Maaskant2010}.

\begin{figure}[!ht]
\centering
\begin{tabular}{cc}
\includegraphics[width=8.5cm]{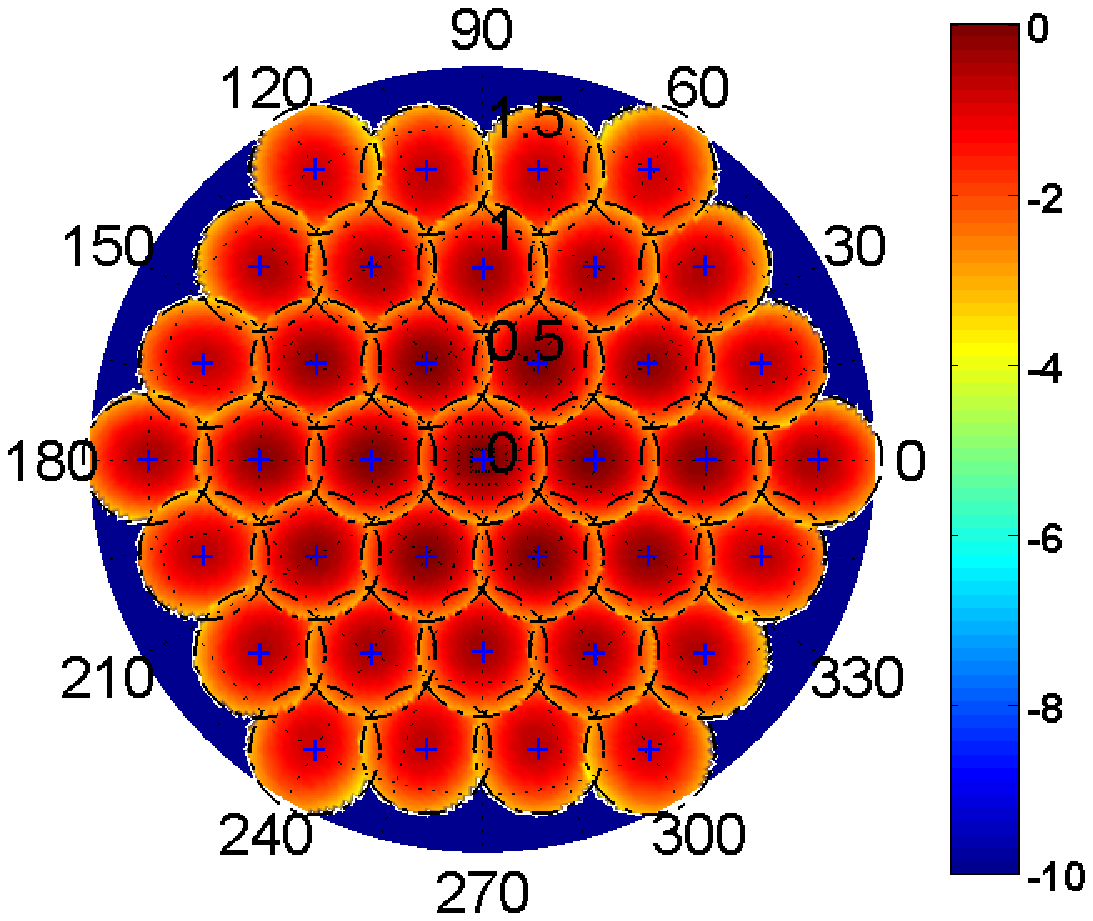}& \includegraphics[width=8.5cm]{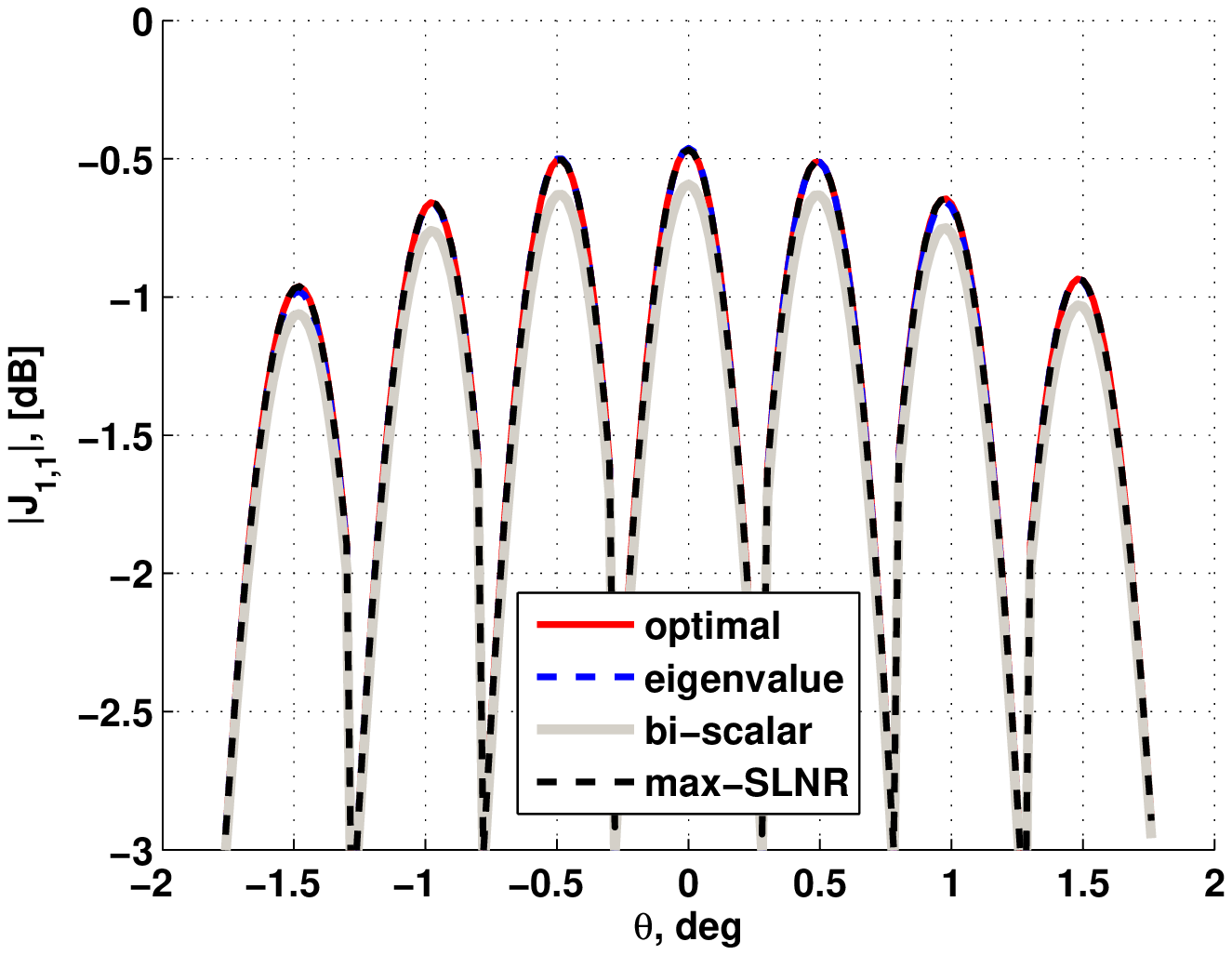}\\
\small (a) $|J_{11}|$ of 37 beams (optimal method) & \small (b) $|J_{11}|$ at $\phi=0^{o}$ for several beamformers\\
\includegraphics[width=8.5cm]{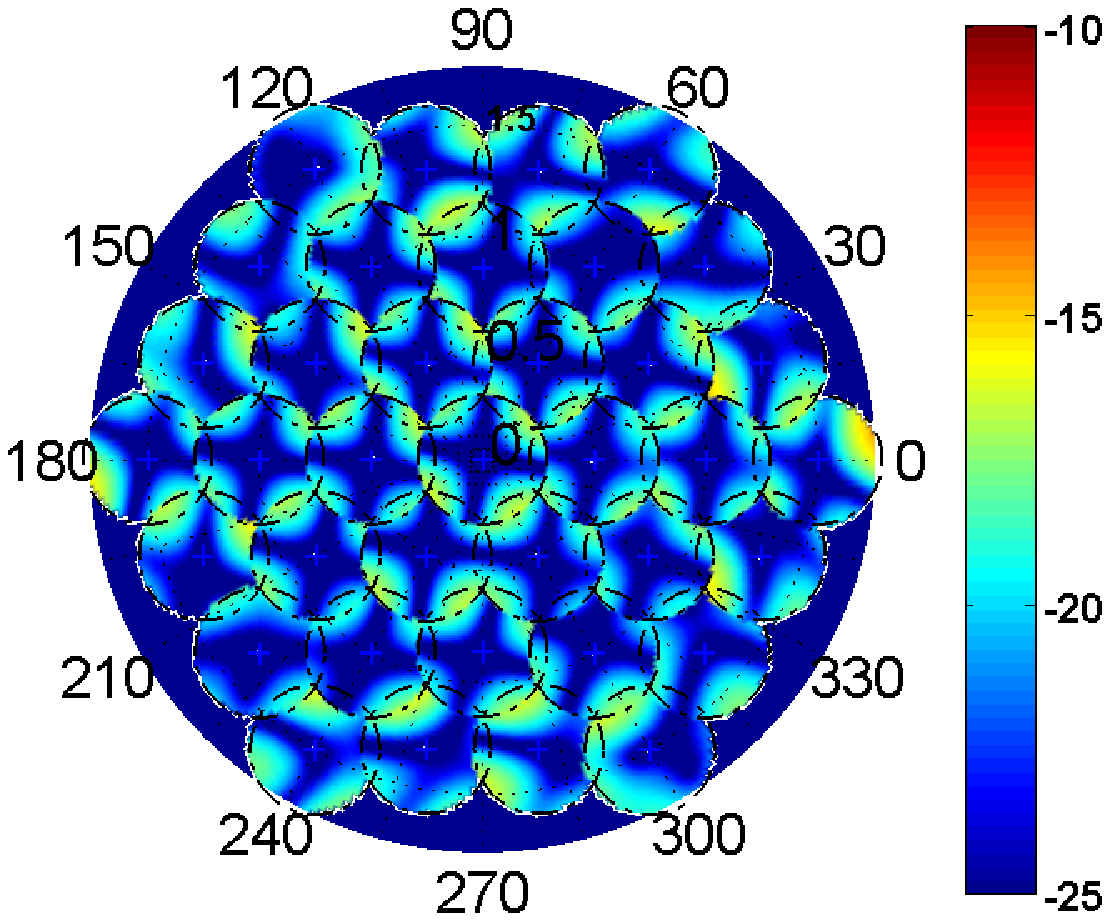}& \includegraphics[width=8.5cm]{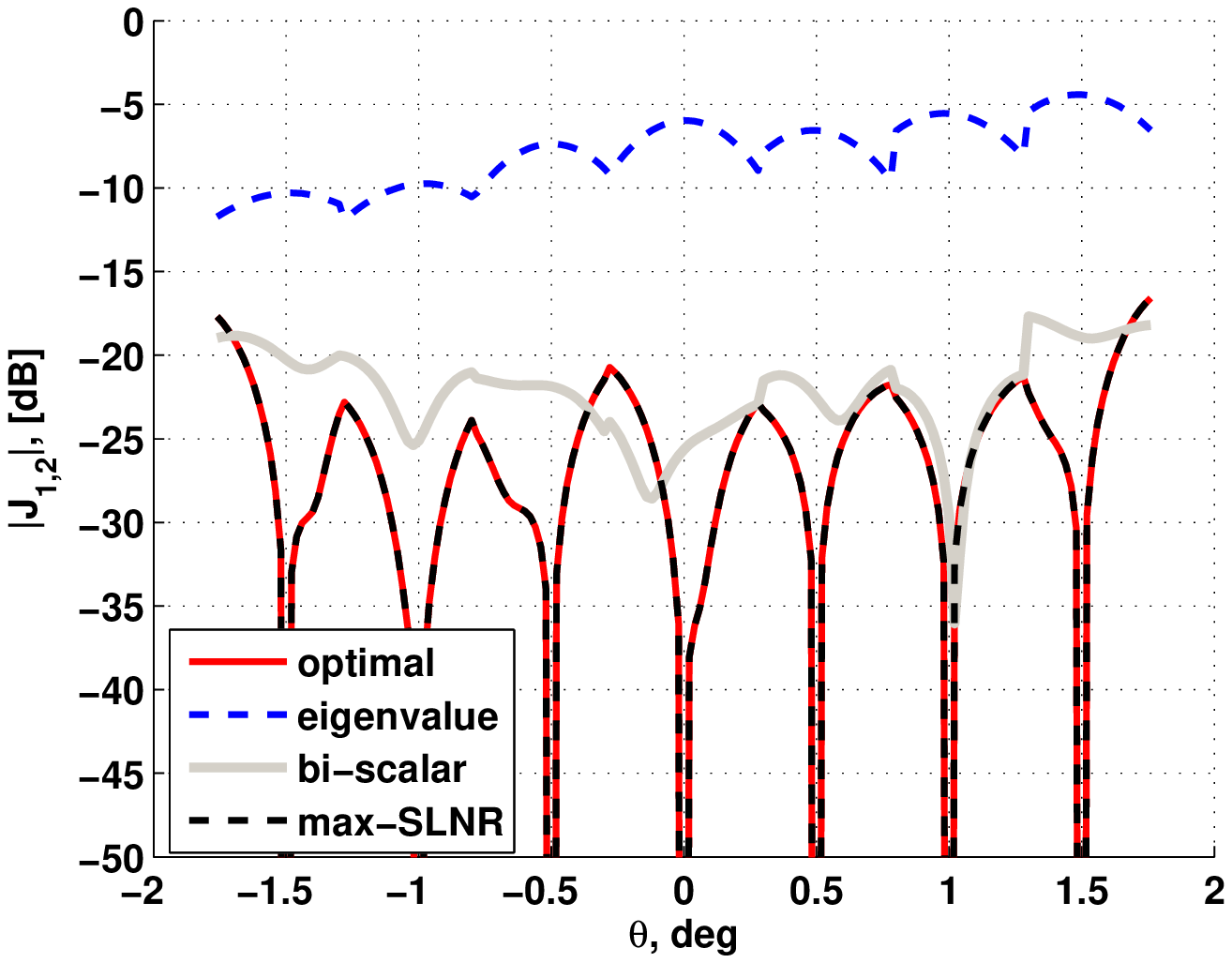}\\
\small (c) $|J_{12}|$ of 37 beams (optimal method) & \small (d) $|J_{12}|$ at $\phi=0^{o}$ for several beamformers \\
\includegraphics[width=8.5cm]{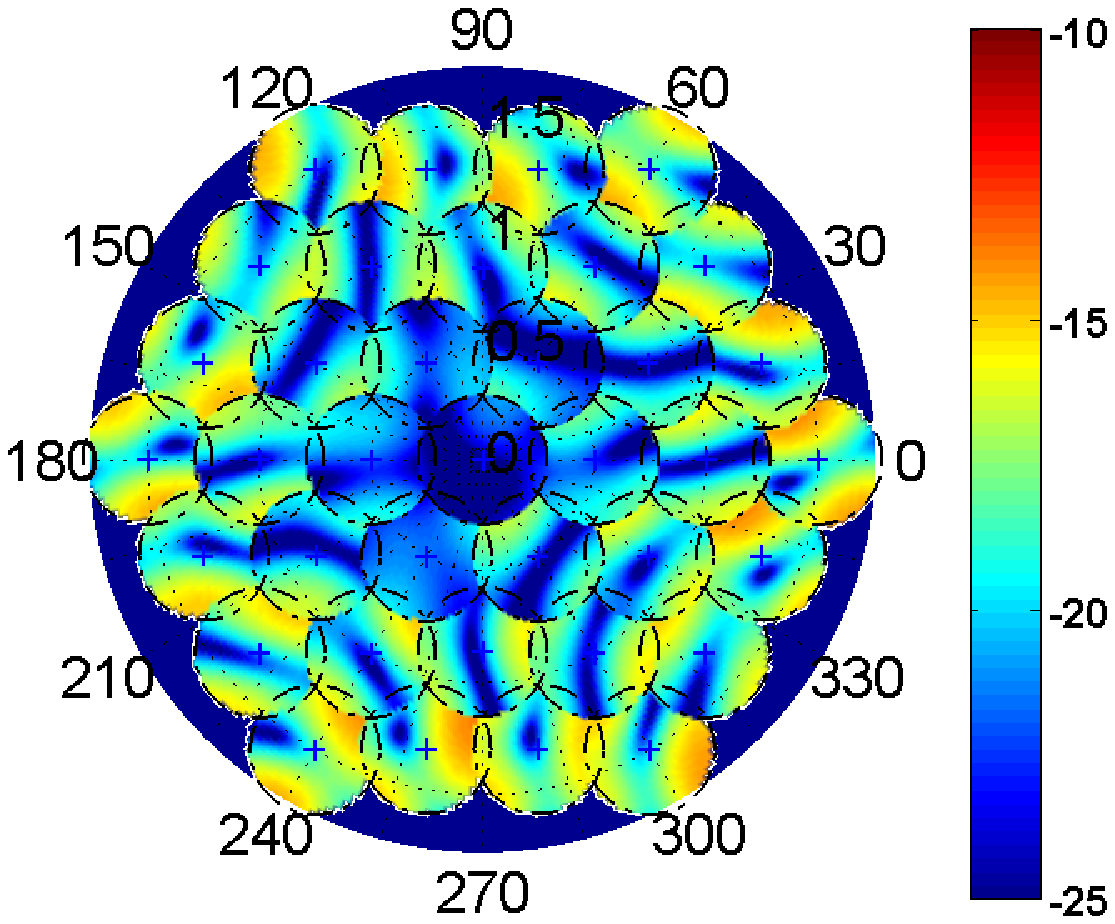}& \includegraphics[width=8.5cm]{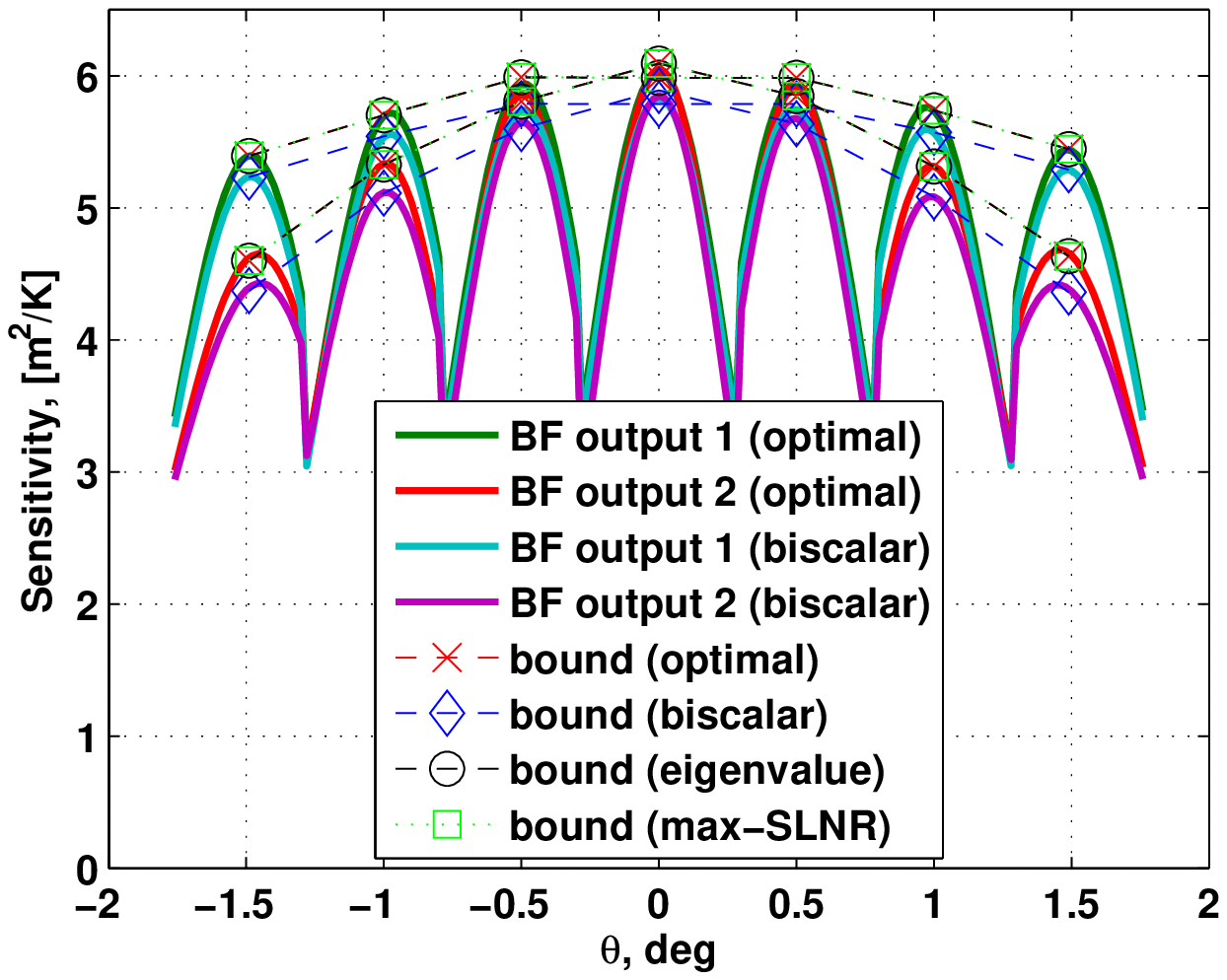}\\
\small (e) $|J_{12}|$ of 37 beams (biscalar method) & \small (f) Beam sensitivities at $\phi=0^{o}$ for several beamformers\\
\end{tabular}
\caption{Simulated Jones matrix elements for several PAF beams and polarimetric beamforming methods. Since the half-power-beamwidth is equal approximately to $0.5^\text{o}$, the beam separation distance is about one half-power beamwidth.}\label{Sensitivity_FOV}
\end{figure}

Figure~\ref{Sensitivity_FOV}(a) illustrates the magnitude of the Jones matrix element $|J_{11}|$ for the optimal method. Element $|J_{11}|$ indicates the beam intensity that is referenced to output 1 of the Jones polarimeter. The results are shown for 37 beams and have been overlayed (not summed) to ease the comparison between the beams within the field of view. Figure~\ref{Sensitivity_FOV}(b) shows $|J_{11}|$ along the $\phi=0^{o}$ direction (for the on-axis and 3x2 off-axis beams) for the four beamforming scenarios. The values of $|J_{11}|$ are the same for the optimal, eigenvalue and max-SLNR methods. The bi-scalar method measures a lower field intensity since half of the antenna output signals are combined into an $x$ beam, and the other half into a $y$ beam, respectively. Figure~\ref{Sensitivity_FOV}(c) plots the Jones matrix element $|J_{12}|$ for 37 beams. The element $|J_{12}|$ measures the degree of cross-polarization leakage for the optimal beamformer. Clearly, the polarization purity is very high in the beam centers, but deteriorates towards the periphery of the beam and can become as high as -15dB for the beams at the edge of the FoV. Figure~\ref{Sensitivity_FOV}(d) shows the cut $\phi=0^{o}$ for $|J_{12}|$ for the four beamforming scenarios. It is remarkable to see that the eigenvector beamformer leads to a cross-polarization leakage which can be as high as -5dB, while the cross-polarization leakage for the beams obtained with the other beamforming schemes is as low as -20dB for the same beam directions. Figure~\ref{Sensitivity_FOV}(e) indicates that for the bi-scalar method, the polarization homogeneity across the beam can be significantly poorer than that of the full-polarization beamformers, and the cross-polarization leakage can rapidly degrade to -12dB for out-of-center beams. Figure~\ref{Sensitivity_FOV}(f) illustrates the beam sensitivities at the two beamformer outputs for the optimal and bi-scalar method. In addition, the sensistivity bounds are shown for the optimal, bi-scalar, eigenvalue, and max-SLNR beamforming methods. These upper and lower bounds are the same for all methods, except for the bi-scalar method, for which these bounds are lower because this scheme does not beamform all antenna output signals and therefore compromises the sensitivity and beam polarization purity, in particular for out-of-center beams.

Clearly, $|J_{1,2}|\rightarrow 0$ at the beam centers for both the optimal and max-SLNR beamformers. Hence, the \emph{IEEE} definitions for cross-polarization, i.e. the cross-polarization discrimination (XPD) and isolation (XPI) values, tend to infinity at the beam centers. This surely meets the SKA specification, which states that the raw cross-polarization level (before calibration) must be lower than -20dB at zenith~\cite{SKAmemo100}. However, the polarimetric homogeneity of the PAF beams is not constant over the FoV and rapidly degrades as a function of off-axis angle of observation, especially for the bi-scalar beamformer. This implies that further calibration must be done to correct for the direction dependent artefacts in interferometric images, in particularly when the dominant source moves away from the centre to the periphery of the beam~\cite{Smirnov2011}.

\bibliography{refs}

\end{document}